\documentclass[fleqn,10pt]{wlscirep}
\usepackage[utf8]{inputenc}
\usepackage[T1]{fontenc}
\usepackage{graphicx}  	
\usepackage{bm}        	
\usepackage{amssymb}   	
\usepackage{amsmath}	
\usepackage{hyperref}
\makeatletter
\newcommand\setcurrentname[1]{\def\@currentlabelname{#1}}
\makeatother

\def\DJ{{\fontencoding{T1}\selectfont\char208}}
\title{Comparative analysis of light storage in antirelaxation-coated and buffer-gas-filled alkali vapor cells}

\author[1]{M.~\DJ uji\'{c}}
\author[1]{D.~Buhin}
\author[1,*]{N.~\v{S}anti\'{c}}
\author[1]{D.~Aumiler}
\author[1]{T.~Ban}

\affil[1]{Institute of Physics, Centre for Advanced Laser Techniques, Bijeni\v{c}ka cesta 46, 10000 Zagreb, Croatia}

\affil[*]{nsantic@ifs.hr}


\begin{abstract}
We explore light storage in antirelaxation-coated and buffer-gas-filled alkali vapor cells, employing electromagnetically induced transparency (EIT) in warm rubidium vapor.
We conduct a comparative study of light storage performance under identical experimental conditions for these two cell types. 
Using a buffer-gas-filled cell resulted in approximately a tenfold improvement in memory efficiency and storage time compared to antirelaxation-coated cells.
Moreover, we demonstrate that memory efficiency can be further enhanced by choosing a near-resonant EIT $\Lambda$-scheme over a resonant one. 
Our findings provide valuable insights for optimizing light storage, thereby contributing to the development of field-deployable quantum memories. 
\end{abstract}
\begin{document}

\flushbottom
\maketitle

\thispagestyle{empty}

\section*{Introduction}

The ability to store and retrieve light pulses on demand has spurred the development of quantum memories (QMs) with perspective applications in quantum communication \cite{sangouard2011}, synchronization of single photon sources \cite{kaneda2017}, production of multiphoton states \cite{nunn2013}, linear optical computing, metrology, magnetometry, and single photon detection \cite{bussieres2013}.   

While systems employing cold atoms \cite{zhao2009, wang2019, saglamyurek2021} and rare-earth atoms \cite{jin2022} demonstrate excellent memory performance characterized by long storage times, high fidelity, and efficiency, there is a growing interest in the scientific community in developing of QMs using warm atomic vapors \cite{eisman2005, reim2011, filkenstein2018, wang2022, buser2022, thomas2019, vurgaftman2013}. 
This increasing interest can be primarily attributed to the simplicity of their technology, making them particularly appealing for real-world applications.
At the core of such quantum memories lies the atomic vapor cell. 
To reduce ground state decoherence resulting from atom-cell wall collisions, vapor cells filled with a buffer gas or vapor cells with an antirelaxation coating deposited on the inner surfaces of the cell can be used.
However, determining the preferred cell type, whether it's a buffer gas cell or one with an antirelaxation coating, poses a challenge due to the absence of comparative studies in the literature that assess the performance of light storage for different cells under identical experimental conditions.
In most prevalent realizations of QMs \cite{wang2022, buser2022}, buffer-gas-filled alkali vapor cells are used, and promising memory performances are demonstrated. 

Motivated by the crucial role of alkali vapor cells in advancing future high-performance, field-deployable quantum memories, we perform a comparative study of light storage in antirelaxation-coated and buffer-gas-filled alkali vapor cells. 
By maintaining identical experimental conditions we ensure similar contributions from various potential decoherence mechanisms, including transit time effects, inelastic atom-atom collisions, residual Doppler broadening, and stray magnetic fields for both cell types. 
As a result, any discernible difference in memory performance can be unequivocally attributed solely to the influence of either the buffer gas or the anti-relaxation coating.
While extensive studies of classical light storage in both antirelaxation-coated \cite{katz2018} and buffer-gas-filled alkali vapor cells \cite{phillips2001, nathaniel2008,hosseini2011} can be found in the literature, and a comprehensive review can be found in \cite{novikova2012}, to our knowledge, there is no comparative study in which the storage performance is measured for both cell types using the same experimental conditions.

Our optical memory is implemented using the electromagnetically induced transparency (EIT) effect in warm rubidium vapor contained in three types of vapor cells: paraffin-coated, alkene-coated, and paraffin-coated cell filled with a few Torr of neon buffer gas. We observe EIT linewidths of a few kHz for the buffer-gas-filled cell and several tens of kHz for the paraffin- and alkene-coated cells. 
Additionally, we employ a theoretical model based on optical Bloch equations which includes the realistic atomic energy level structure, i.e., all hyperfine levels. The results of the theoretical model agree very well with the EIT experiment, which indicates that the widely-used simplified three-level approximation is not sufficient for a realistic analysis even in the case of classical light. This finding is even more significant for the theoretical treatment of QMs, given that quantum states are very fragile and a real analysis of the effects of noise on the storage of quantum states is required. 
The best light storage performance is achieved in the neon-filled gas cell, where in the near-resonant EIT condition, we measure a storage efficiency of $25\%$ and a storage time of up to 1.2 ms.

A detailed description of our experimental setup and procedures can be found in Section \nameref{experiment}.
In Section \nameref{EITspectra}, we provide measurements of EIT spectra in both resonant and non-resonant schemes, alongside a discussion of basic concepts essential for gaining further insight into the work.
Moving forward to Section \nameref{CompEIT}, we provide EIT spectra measured for each of the three cell types, i.e. paraffin-coated, alkene-coated, and buffer-gas-filled cells. 
Subsequently, we quantify the storage time and memory efficiency for each cell, all under the same experimental conditions, as detailed in Section \nameref{MemoryComp}. 
These conditions encompass matching EIT linewidth, shape and duration of the input probe pulse, rubidium vapor optical density, probe and coupling laser frequencies, as well as the measurement protocol itself. 
Finally, in Section \nameref{Optimal}, we investigate the temperature dependence of storage time and efficiency for the cell exhibiting the best storage performance, particularly under near-resonant EIT conditions.

\section*{Results}

\subsection*{Experimental setup} 
\setcurrentname{Experimental setup}
\label{experiment}

\begin{figure}[h]
\centering
\includegraphics[width=0.7\textwidth]{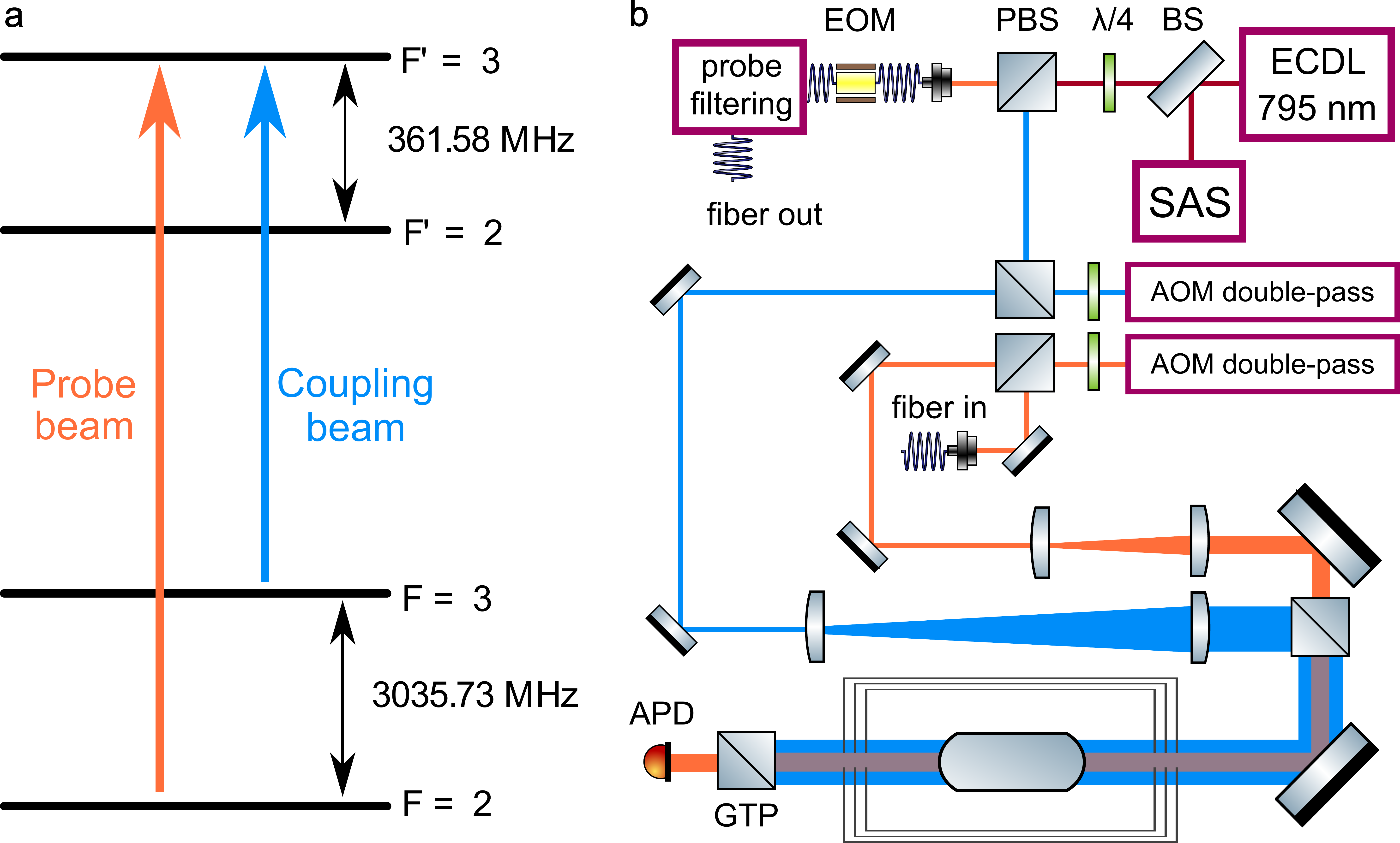}
\caption{(a) Energy level scheme used in the experiment. A strong coupling field excites the $|5S_{1/2};F=3\rangle - |5P_{1/2};F'=3\rangle$ transition, while a weak probe field excites the $|5S_{1/2};F=2\rangle - |5P_{1/2};F'=3\rangle$ transition. (b) Experimental setup. The single external cavity diode laser (ECDL) used in the experiment is stabilized using saturated absorption spectroscopy (SAS). Its output beam is divided into two orthogonally polarized beams using a polarizing beam splitter (PBS). The probe beam is fed into a fiber-coupled electro-optical modulator (EOM), after which the blue sideband, corresponding to the $|5S_{1/2};F=2\rangle - |5P_{1/2};F'=3\rangle$ transition, is selected using an optical filtering cavity. Both the probe and the coupling beam are then shifted in frequency using acousto-optical modulators (AOM) in a double-pass configuration, after which they undergo expansion via telescopes and recombination using a PBS. The combined beams pass through the magnetically isolated cell, after which a Glan-Thompson polarizer (GTP) is used to remove the coupling light. The remaining light is then detected by an avalanche photodiode (APD). BS - beam splitter, $\lambda$/4 - quarter-wave plate.
	}
	\label{Fig1}
\end{figure}

Our optical memory is based on EIT in the hyperfine $\Lambda$ scheme in warm $^{85}$Rb atoms, see Fig.~\ref{Fig1} for relevant energy levels and an optical layout of the experiment. 

In our study, we utilized a set of three cylindrical quartz cells, each with a length of 75 mm and a diameter of 25 mm, containing isotopically enriched $^{85}$Rb atoms. Two of these cells were evacuated and had their inner walls coated with paraffin and alkene antirelaxation coating, respectively. The third cell was paraffin-coated and filled with 5 Torr of Ne buffer gas.
All cells were purchased from Precision Glassblowing and were made according to our specifications.
To minimize the influence of stray magnetic fields, we encased the rubidium cell within three layers of mu-metal magnetic shielding. 
The cell's temperature can be adjusted from room temperature to 60$^{o}$C.
This is achieved by heating two aluminum blocks between which the cell is placed. These blocks are heated externally by copper heat pipes running from outside the mu-metal shielding. 
This temperature control allowed us to modulate the rubidium vapor optical depth (OD), from OD=0.75 at room temperature to OD=6 at 45~$^{o}$C.

We initiate the EIT measurements by setting the coupling laser frequency using an AOM.
Next, we then scan the probe laser frequency with a separate AOM around the two-photon resonance while simultaneously measuring the probe transmission.
For a given set of experimental parameters, we record three key measurements: the EIT transmission spectrum ($I$), the off-resonant probe transmission ($I_0$), and the background signal ($B$).
The off-resonant probe light is obtained by transmitting the red-shifted sideband instead of the blue one through the filtering cavity. 
Meanwhile, the background signal is acquired with the probe light turned off, and is the result of the coupling beam leakage through the Glan-Thompson polarizer to the probe detector.
To calculate the optical depth for the probe light, which represents the EIT absorption spectrum, we apply the Beer-Lambert law, expressed as $OD = -\ln\left[(I - B)/(I_0 - B)\right]$.
We perform this measurement protocol 30 times under identical experimental conditions and subsequently calculate the average to obtain the final EIT spectrum.
\textbf{In the case of resonant EIT}, we adjust the coupling laser to the $|5S_{1/2};F=3\rangle \rightarrow |5P_{1/2};F'=3\rangle$ transition frequency.
The probe intensity is maintained at a constant level of 2.8 $\mu$W/cm$^{2}$, and the intensity of the coupling beam is varied between 0.15 and 2.5~mW/cm$^{2}$, corresponding to probe and coupling Rabi frequencies of 2$\pi\cdot$~50.6 kHz and 2$\pi\cdot$~0.5 to 2$\pi\cdot$~1.5 MHz, respectively.
\textbf{In the case of near-resonant EIT}, we vary the detuning of the coupling laser from the $|5S_{1/2};F=3\rangle - |5P_{1/2};F'=3\rangle$ transition, denoted as $\Delta_{c}$, within a range of -1000 to 1000 MHz. To achieve these detuned frequencies we no longer use saturation absorption spectroscopy to stabilize the ECDL output frequency, but rather stabilize it using a wavelength meter.

For the implementation of an optical memory, i.e., light storage and retrieval, we apply a commonly used protocol \cite{phillips2001, novikova2012, phillips2008}.
The coupling and the probe lasers have fixed frequencies that satisfy the two-photon EIT resonance.
First, we turn on the coupling beam, after which we apply an exponential probe beam pulse with a duration of 10~$\mu$s, which is optimized for the highest memory efficiency using a protocol described in Ref. \cite{novikova2007}. 
At the peak of the probe pulse, we shut off the probe and coupling beams simultaneously.
After waiting for a variable storage time, we turn on the coupling beam and monitor the intensity of the retrieved probe pulse with an avalanche photodiode placed after the rubidium cell, see Fig.~\ref{Fig1}. 
The shaping of the probe and coupling pulses as well as the precise time sequence of the experimental protocol is achieved using the probe and coupling beams AOMs controlled by a common two-channel RF synthesizer. 
We repeat the measurement protocol 1000 times for identical experimental conditions, and average the measured traces to obtain the retrieved pulse for a given storage time. Finally, we calculate the storage efficiency as the ratio of the retrieved and incident pulse areas.

\subsection*{EIT spectra}
\setcurrentname{EIT spectra}
\label{EITspectra}

\begin{figure}[h]
\centering
	\includegraphics[width=0.95\textwidth]{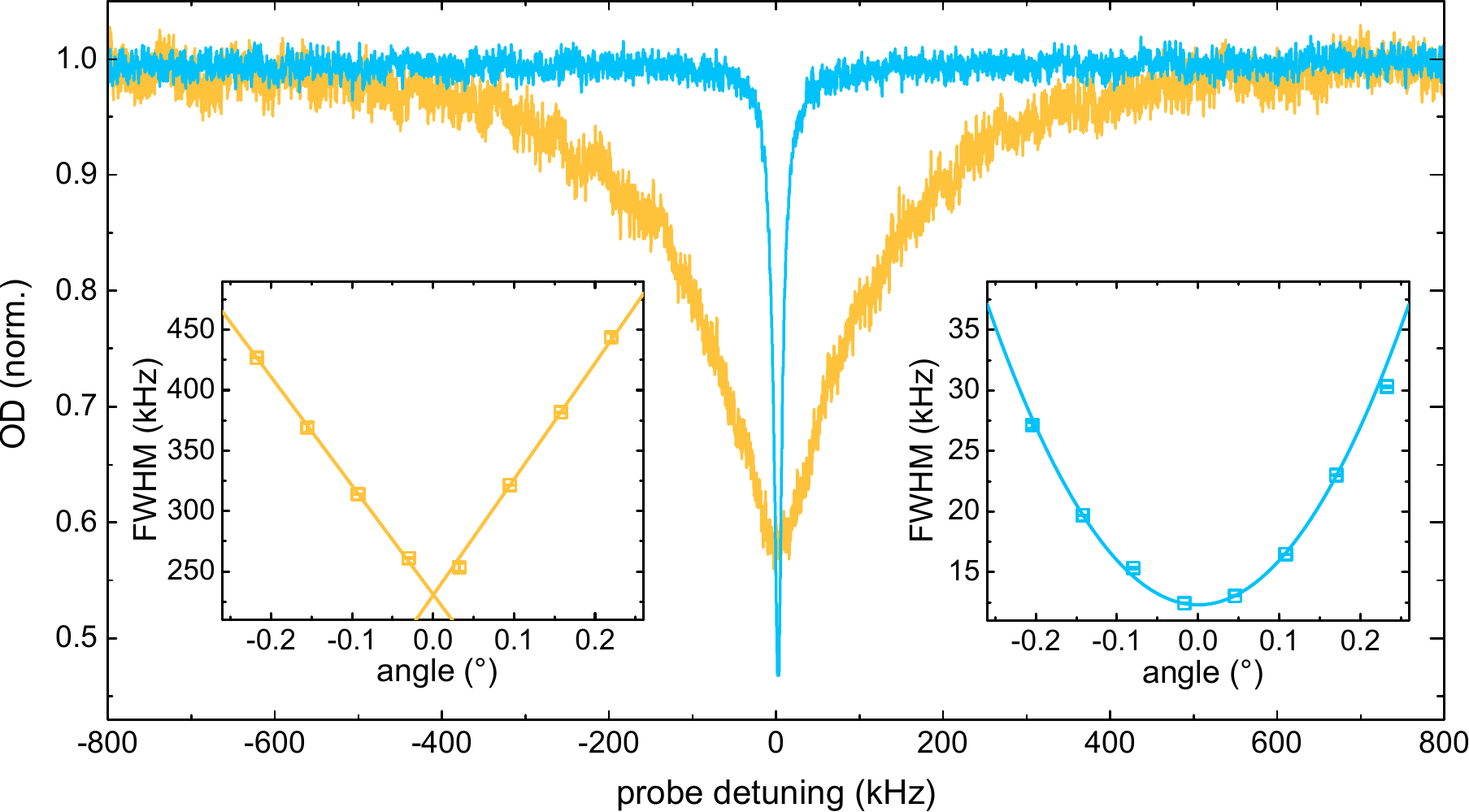}
	\caption{EIT spectra measured for the reference cell (yellow) and the neon-filled cell (blue). Insets depict the dependence of the EIT spectrum linewidth on the angle between the probe beam and the coupling beam for the reference cell (left) and the neon-filled cell (right). 
	}
	\label{Fig2}
\end{figure}

In Fig.~\ref{Fig2}, we present EIT spectra measured for the buffer-gas-filled cell (blue) and a reference cell (yellow) that lacks an anti-relaxation coating on the walls and contains no buffer gas. 
These measurements are conducted using room temperature rubidium vapor under the resonant EIT condition, with a coupling beam intensity of 0.29 mW/cm$^{2}$.

The EIT spectrum acquired in the reference cell exhibits a cusp-like lineshape described by the function e$^{-|\Delta_{p}|t_{TT}}$, where $|\Delta_{p}|$ represents the probe detuning, and $t_{TT}$ is the transit time \cite{filkenstein2023, thomas1980}. 
By fitting this exponential function to the measured EIT spectra, we determine a transit time of 7.2 $\mu$s for the cell with no anti-relaxation coating and no buffer gas, corresponding to a full width at half maximum (FWHM) of approximately
190 kHz.
Transit time, indicates the average interaction time of atoms with the probe light field, results from the transverse motion of atoms within the finite-sized probe beam. 
With no anti-relaxation coating and no buffer gas to reduce the transit time, its dominance is understandable, as other broadening mechanisms stemming from magnetic field inhomogeneities, power broadening, and atomic collisions are on the order of a few kHz or less, see Sec. \nameref{CompEIT}. 

The residual Doppler broadening, given by $\Gamma_{D,res}=|\vec{k_{c}}-\vec{k_{p}}|v_{th}$, where $\vec{k_{c}}$ and $\vec{k_{p}}$ are the coupling and probe wave vectors, respectively, is expected to be $\approx2\pi\cdot$~1.7~kHz in the case of perfect beam overlap, i.e., for $\theta= 0$, where $\theta$ represents the angle between the probe and coupling beams, and $v_{th}$ = $\sqrt{kT/m}$ = 170~m/s is the RMS thermal velocity.
Increasing the $\theta$ angle results in an additional increase of the residual Doppler broadening.
In evacuated cells, for a $\Lambda$-scheme with close-lying coupling and probe frequencies, and small $\theta$, the residual Doppler broadening is proportional to $\theta$ \cite{filkenstein2023, derose2019}. 
The left inset of Fig.~\ref{Fig2} illustrates this relationship by showing measured EIT linewidths as a function of $\theta$ (symbols) and the linear fits to the data (lines).
In our experiment, we are able to almost completely cancel out additional residual Doppler broadening caused by $\theta$, i.e., to work close to $\theta$ zero, by coupling both beams into the same optical fiber after the cell.

To increase the interaction time and consequently reduce transit time broadening both antirelaxation-coated and buffer-gas-filled cells are used.
In antirelaxation-coated cells, the inner walls are coated with a material designed to minimize decoherence-inducing collisions. 
This coating significantly increases the likelihood of atoms retaining coherence upon contact with the cell wall, and can consequently traverse the probe beam multiple times before decoherence occurs.
Alternatively, a buffer gas is introduced into the vapor cell to induce velocity-changing collisions that prevent atoms from exiting the probe beam. 
In the case of a buffer-gas-filled cell, the resonant EIT spectrum approximates a Lorentzian distribution, and a reduction in EIT linewidth by a factor of 14.5 compared to the reference cell is observed, as shown in Fig.~\ref{Fig2}.
The reduction in the EIT linewidth originates from the frequent velocity-changing collisions with the buffer gas, which increase the transit time, thereby decreasing the transit time broadening, and induce Dicke narrowing.
For our experimental conditions, we calculated transit time of 45 ms corresponding to the transit time broadening of 354 Hz.
This calculation is performed using the relation (2) in Ref. \cite{arimondo1996}, with $D$=30 cm$^{2}$/s calculated using data from the Ref. \cite{chrapkiewicz2014}, a pressure of Ne gas $p$=5 Torr, a probe beam radius of $R$=2.8 mm, and a mean free path of $\lambda$=4 $\mu$m obtained from the Ref. \cite{shuker2007}. 
Due to Dicke narrowing, residual Doppler broadening is proportional to $\theta^{2}$ \cite{filkenstein2023, derose2019}, making it much less sensitive to the alignment of the coupling and probe beams. 
The right inset of Fig.~\ref{Fig2} illustrates this by showing Dicke-narrowed EIT linewidths as a function of $\theta$ measured for rubidium vapor contained in the buffer-gas-filled cell (symbols) and a quadratic fit to the data (line).
It should be noted here that the EIT spectra shown in Fig.~\ref{Fig2}, and all spectra related to the data in the insets, are measured using a coupling beam intensity of 0.29 mW/cm$^{2}$, implying that for $\theta$=0, they are limited by the residual Doppler broadening along with power broadening.

\begin{figure}[h]
\centering
\includegraphics[width=0.95\textwidth]{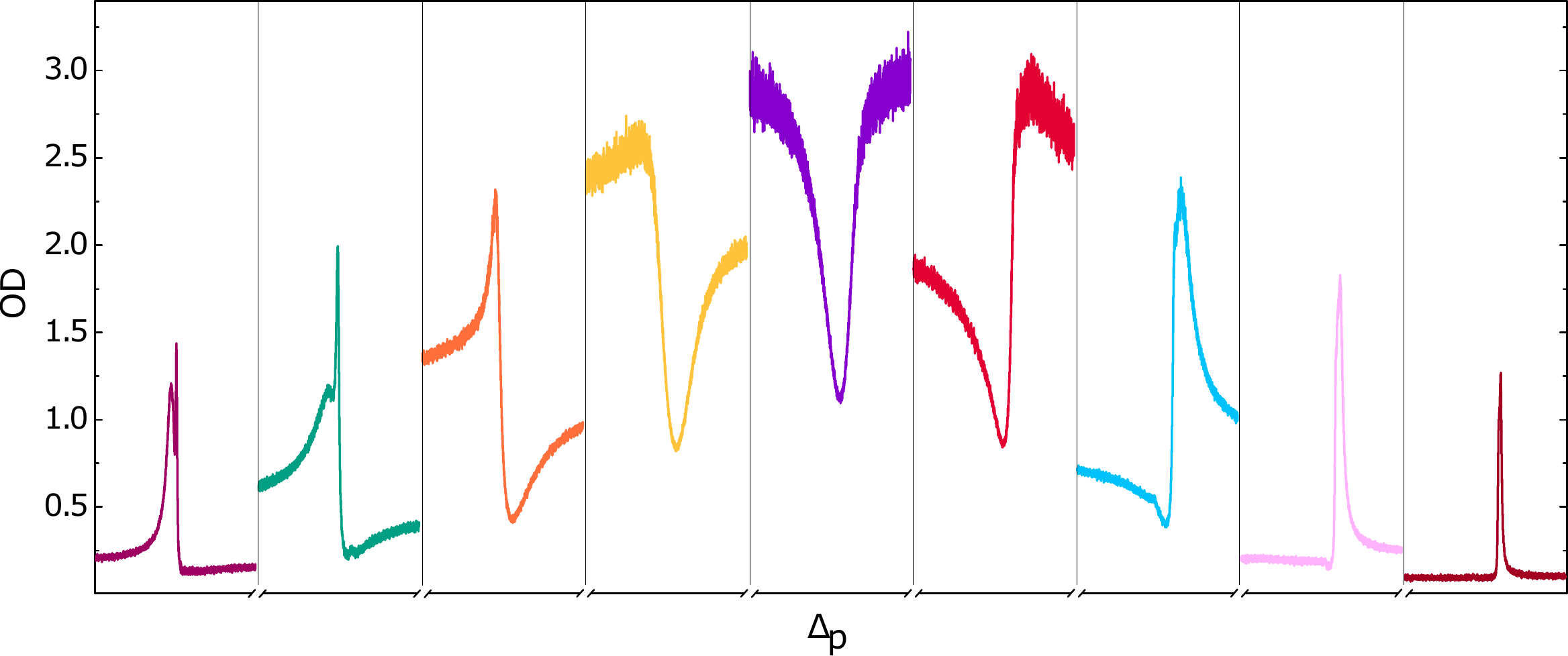} 
\caption{EIT spectra for different coupling beam detunings in the neon-filled cell at 33 °C. Each spectrum was measured by scanning the probe laser frequency across the two-photon resonance in a 500~kHz range. Coupling beam detuning is from -800 MHz to 800 MHz in steps of 200 MHz, when going from the left to the right panel. 
	}
	\label{Fig3}
\end{figure}

In Fig.~\ref{Fig3}, we present EIT spectra obtained for different values of $\Delta_{c}$, using room temperature rubidium vapor within the buffer-gas-filled cell, with a coupling beam intensity of 2.5 mW/cm$^{2}$.
The EIT resonance is symmetric when the coupling laser is precisely tuned to the $|5S_{1/2};F=3\rangle - |5P_{1/2};F'=3\rangle$ optical transition, and begins to show asymmetry as the coupling laser detuning increases.
Upon further increase or decrease of the coupling laser frequency beyond the natural linewidth of $\Gamma=2\pi$ $\cdot$ 5.75 MHz \cite{steck2}, the EIT feature initially shows not only a decrease in absorption, but also an increase. Eventually, it fully transforms into an absorption resonance.

For a given $|\Delta_{c}|$, the near-resonant EIT spectra exhibit opposite symmetry for the blue and red-detuned coupling beams. In other words, when the coupling beam is red-detuned, the EIT resonance occurs at a higher probe frequency than the absorption resonance, and vice versa for the blue-detuned coupling beam.
Furthermore, the complete transformation of the EIT feature into an absorption resonance occurs at smaller $|\Delta_{c}|$ values when using a blue-detuned coupling beam. This behavior is attributed to the contribution of the fourth hyperfine level $|5P_{1/2};F'=2\rangle$ to the EIT signal when the coupling beam is red-detuned from the $|5S_{1/2};F=3\rangle - |5P_{1/2};F'=3\rangle$ transition, as shown in Fig.~\ref{Fig1}(a).

\subsection*{Comparative study of EIT in antirelaxation-coated and buffer-gas-filled alkali vapor cells}
\setcurrentname{Comparative study of EIT}
\label{CompEIT}

Light storage performance in vapor cells is closely related to the EIT linewidth and contrast.
Indeed, for EIT optical memories based on a $\Lambda$ scheme, the EIT linewidth contains information about the spin coherence of the ground hyperfine levels which is directly related to the memory storage time.
Therefore, in order to benchmark the light storage performance, we first measure the EIT linewidths for all three cells (paraffin-coated, alkene-coated, and parafin-coated Ne buffer-gas-filled) in identical experimental conditions. 

\begin{figure*}
\centering
	\includegraphics[width=0.95\textwidth]{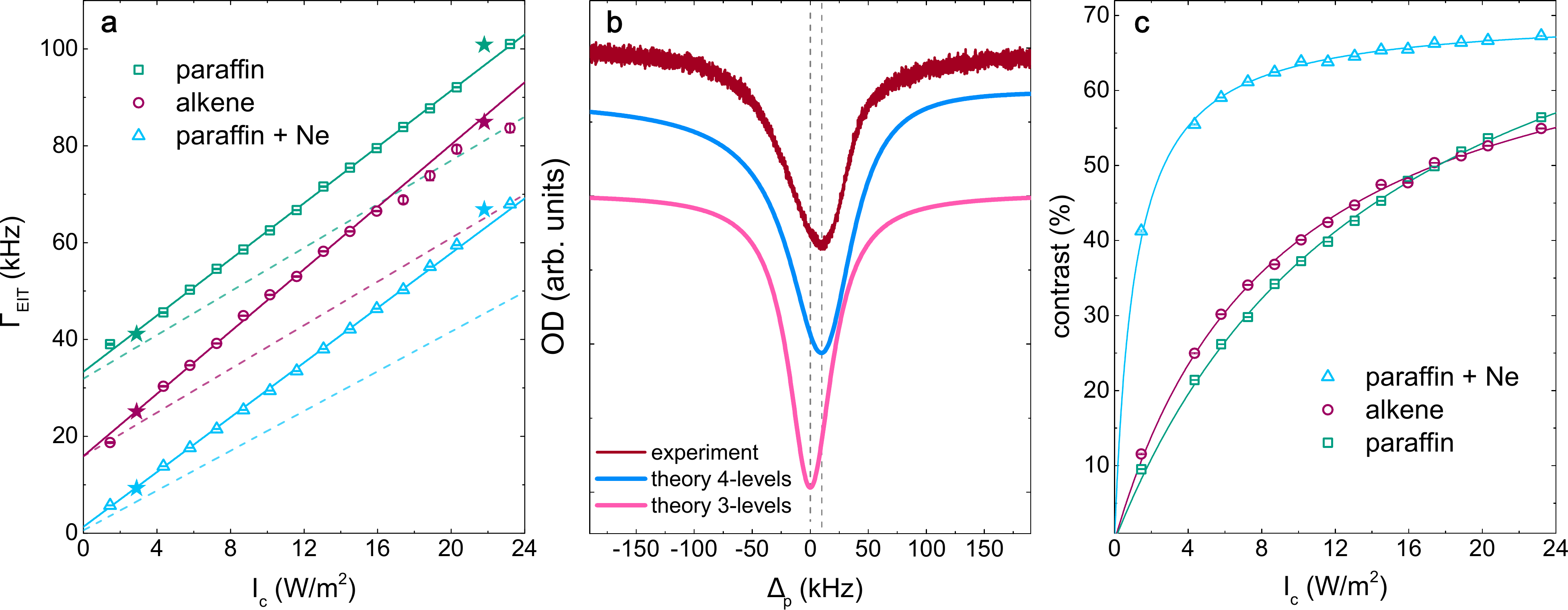}
	\caption{(a) EIT linewidths as a function of the coupling light intensity for all three cells. Solid lines represent linear fits to data, dashed lines represent calculated EIT linewidths using 3-level atom model, whereas the star symbols represent calculated EIT linewidths using a 4-level atom model. (b) Comparison of the calculated EIT spectra obtained using a 3- and 4- level atom model with the measured spectrum. (c) EIT contrast as a function of the coupling light intensity for all three cells.   
	}
	\label{Fig5}
\end{figure*}

In Fig.~\ref{Fig5}(a), we present EIT linewidths as a function of the coupling light intensity for all three cells, all measured at room temperature using $^{85}$Rb vapor.
For the lowest coupling laser intensity, we observe EIT linewidths ranging from a few kHz (Ne-filled cell) to tens of kHz (alkene- and paraffin-coated cells).

Various broadening mechanisms contribute to the EIT linewidth, including magnetic field inhomogenities, power broadening, atomic collisions, atomic transit-time, and residual Doppler broadening \cite{filkenstein2023, derose2023}.
We can disregard the influence of magnetic field non-uniformities since we employ a hyperfine $\Lambda$ scheme within a magnetically shielded cell.
Atomic collision broadening resulting from Rb-Rb atom collisions is on the order of a few Hz \cite{harper1972, vanier1998}, making it negligible in comparison to the other broadening mechanisms.
The contribution of power broadening is evident in Fig.~\ref{Fig5}(a), as the EIT linewidth increases linearly with the coupling laser intensity. 
To resolve the effect of power broadening, we fit a linear function (solid line in Fig.~\ref{Fig5}(a)) to the measured EIT linewidths (represented by squares, circles, and triangles in Fig.~\ref{Fig5}(a)). By examining the line intercepts, we can extract the EIT linewidths limited only by transit-time broadening and residual Doppler broadening. For perfectly co-propagated coupling and probe beams, the residual Doppler broadening contribution is 1.7~kHz, as discussed in Sec. \nameref{EITspectra}.
The line intercepts yield values of $b = (1.5 \pm 0.4)$ kHz for the Ne-filled cell, $b = (16 \pm 1)$ kHz for the alkene-coated cell, and $b = (33.3 \pm 0.4)$ kHz for the paraffin-coated cell. 
These results indicate that residual Doppler broadening is the primary contributing mechanism in the Ne-filled cell, while transit-time broadening dominates in the anti-relaxation-coated cells.
Different intercepts for the anti-relaxation-coated cells come as a result of different molecular structures of the coating materials, namely alkene and paraffin. 
Based on our findings, it can be concluded that a cell coated with alkene will experience approximately twice as many „good“ collisions at the wall, i.e., collisions in which decoherence does not occur, compared to a cell coated with paraffin.

The linear dependence of the EIT linewidth on the coupling laser intensity is theoretically predicted in a three-level $\Lambda$ system under the weak probe approximation \cite{filkenstein2023, derose2023}. This relationship directly links the intercept to the decoherence rate of the ground spin states. 
However, as seen in Fig.~\ref{Fig5}(a,b), the simplified three-level approximation may not be sufficient to describe a realistic experiment where additional energy levels contribute to the EIT signal. 
This is the case with $^{85}$Rb atoms, where the $|5P_{1/2};F=2\rangle$ hyperfine level is situated approximately 362 MHz below the $|5P_{1/2};F=3\rangle$ level, as shown in Fig.~\ref{Fig1}(a).
In such a system, two EIT lambda schemes are simultaneously realized, each corresponding to a different $|5P_{1/2};F=2,3\rangle$ excited hyperfine level. This increases the complexity of the atom-light interaction and influences the final atomic populations and coherences. A similar behavior can be anticipated in the case of the $^{87}$Rb atom. However, owing to the energy separation of $\approx$816 MHz, the effect is somewhat smaller, i.e., the EIT spectrum should more closely resembles the 3-level spectrum.

To theoretically investigate the EIT signals for a realistic $\Lambda$ system, we employ standard density matrix formalism in a four-level $^{85}$Rb atom interacting with two continuous-wave (cw) laser fields.
For further details regarding the model, please refer to the Sec. \nameref{methods}.
For fixed $\Delta_c$ = 0, $E_p$ = 3 V/m and a given $E_c$, we calculated the probe absorption coefficient, $\alpha_{p}$, for different probe field detunings, $\Delta_p$.  
From the calculated absorption spectra, we obtained the EIT FWHM and compared it to the experiment.
The agreement between the measured EIT linewidths and the calculated ones is excellent.
For enhanced clarity, Fig.~\ref{Fig5}(a) displays calculated EIT linewidths for only two coupling electric fields (stars), specifically, $E_c$ = 23 V/m and $E_c$ = 64 V/m, corresponding to 2.9 W/m$^{2}$ and 21.8 W/m$^{2}$, where the intensity is calculated using the relation $I=\frac{1}{2}cn\epsilon_o{E}^2$, with $c$ equal to the speed of light, $n$ the index of refraction, and $\epsilon_o$ the permittivity of vacuum.
The same figure also shows the EIT linewidths calculated with a simplified three-level model.
These calculations employ the same parameters as the four-level model, except that $\mu_{13} = \mu_{23}$ = 0.
The results indicate that with increasing coupling light intensity, a significant discrepancy arises between the three-level theory and the experimental data.

The failure of the three-level model to reproduce the measured EIT signals is further supported by Fig.~\ref{Fig5}(b) where the measured (red) and calculated EIT spectra are presented as a function of probe detuning.
Consistent with predictions from the literature \cite{derose2023}, the probe absorption calculated using the three-level model (violet) takes the form of a Lorentzian centered at $\Delta_p = 0$.
However, the measured EIT spectrum takes on an asymmetric lineshape with its center shifted to larger probe detuning, a behavior that is well reproduced by the four-level model (blue).

In Fig.~\ref{Fig5}(c) we show measured EIT contrasts, or amplitude of the transparency, (symbols) as a function of the coupling light intensity for all three cells measured at room temperature $^{85}$Rb vapor. 
The solid curve is a fit to the measured data of the form $I_{c} / (a+I_{c})$, where $a$ is a free parameter.
The functional form is in accordance with the theory, however, the measured contrast is lower than predicted.
This discrepancy can be attributed to the reduction of the steady-state atoms in the dark state due to different dephasing mechanisms, such as population leakage, and absorption \cite{derose2023}.
The Ne-filled cell exhibited the highest contrast, approximately 68$\%$, while the alkene- and paraffin-coated cells showed similar contrasts, reaching up to 50$\%$.  

\subsection*{Comparative study of light storage in antirelaxation-coated and buffer-gas-filled alkali vapor cells}
\setcurrentname{Comparative study of light storage}
\label{MemoryComp}

\begin{figure}[h] 
\centering
\includegraphics[width=0.7\textwidth]{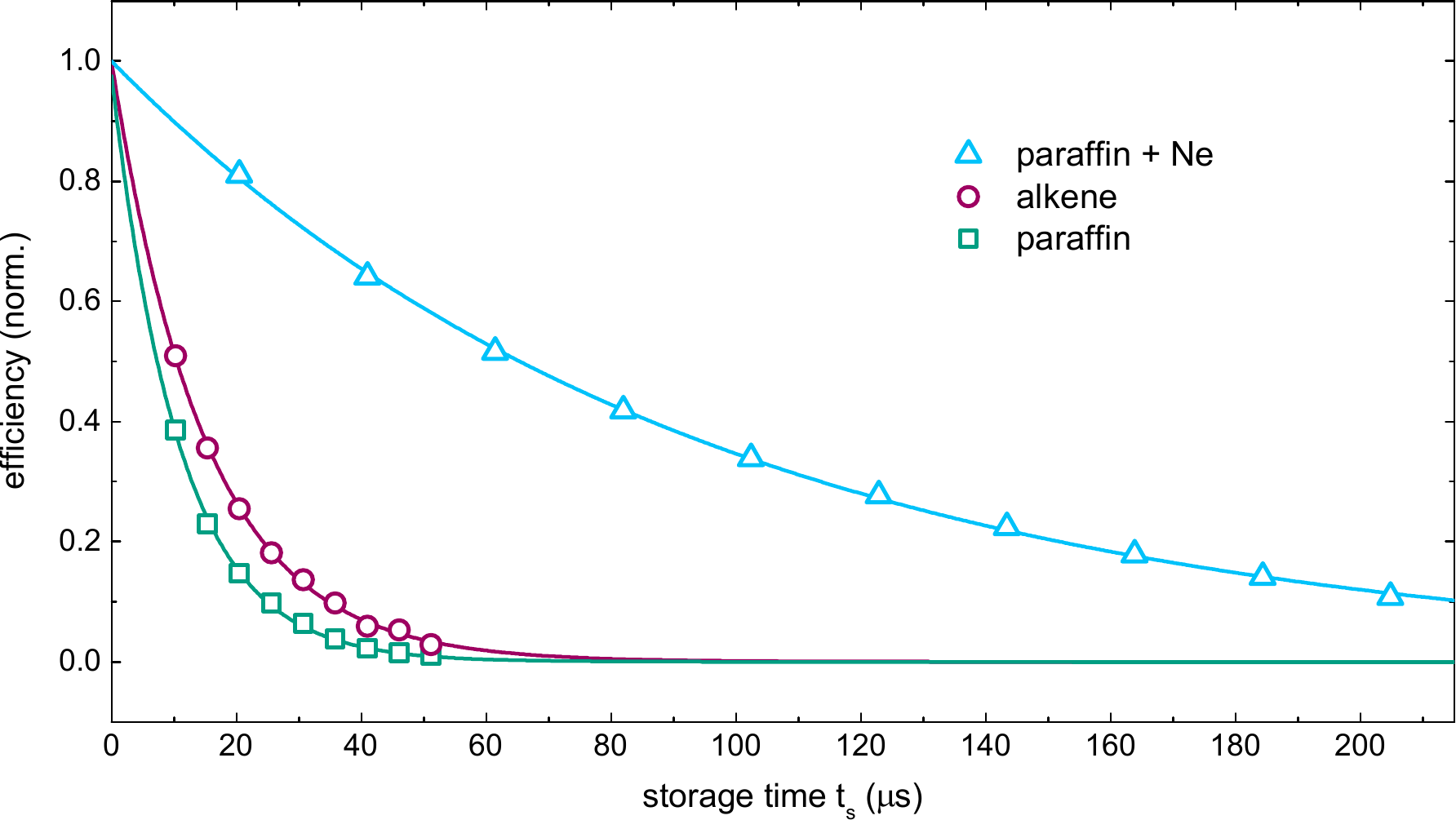}
\caption{Normalized optical memory efficiency as a function of storage time for all three cells. Solid lines are exponential decay fits to the data. We obtained values of $\tau$ = (95 $\pm$ 3) $\mu$s for the Ne-filled cell, $\tau$ = (14.9 $\pm$ 0.9) $\mu$s for the alkene-coated cell, and $\tau$ = (10.3 $\pm$ 0.6) $\mu$s for the paraffin-coated cell.   
	}
	\label{Fig6}
\end{figure}

Light storage is achieved using the optical memory protocol described in Sec. \nameref{experiment}.
In Fig.~\ref{Fig6} we show the measured normalized storage efficiency as a function of the storage time for paraffin-coated, alkene-coated, and Ne-filled cells.
The measurements were performed for $^{85}$Rb vapor at 33~$^{o}$C, resonant EIT ($\Delta_c=\Delta_p = 0$) with a FWHM of 100 kHz.
We fit an exponential decay function $e^{-t/\tau}$ to the measured data, where $\tau$ denotes the memory lifetime, and find $\tau$ = (95 $\pm$ 3) $\mu$s for the Ne-filled cell, $\tau$ = (14.9 $\pm$ 0.9) $\mu$s for the alkene-coated cell, and $\tau$ = (10.3 $\pm$ 0.6) $\mu$s for the paraffin-coated cell.
The obtained memory lifetimes agree well with the decoherence times of ground hyperfine levels, calculated as $1 / (2\pi\cdot\gamma_{12}^{coll})$.
The shorter memory lifetime observed in the the Ne-filled cell with respect to the lifetime expected from the measured EIT linewidth suggests that the initial coherent ground state preparation may not be optimized for short probe pulses.
The measured memory efficiencies are relatively low for all three cells, $\approx1.3\%$ for the Ne-filled cell and $\approx1\%$ for alkene- and paraffin- coated cells.

\subsection*{Optimal light storage in buffer-gas-filled alkali vapor cell}
\setcurrentname{Optimal light storage}
\label{Optimal}

From the previously presented results, it is evident that the Ne-filled cell shows better memory performance than the antirelaxation-coated cells.
Therefore, going further, we limit our investigation to the Ne-filled cell memory and optimize its performance by increasing the temperature of the rubidium vapor, and using a near-resonant EIT $\Lambda$-scheme.

\begin{figure}[h]
\centering
\includegraphics[width=0.95\textwidth]{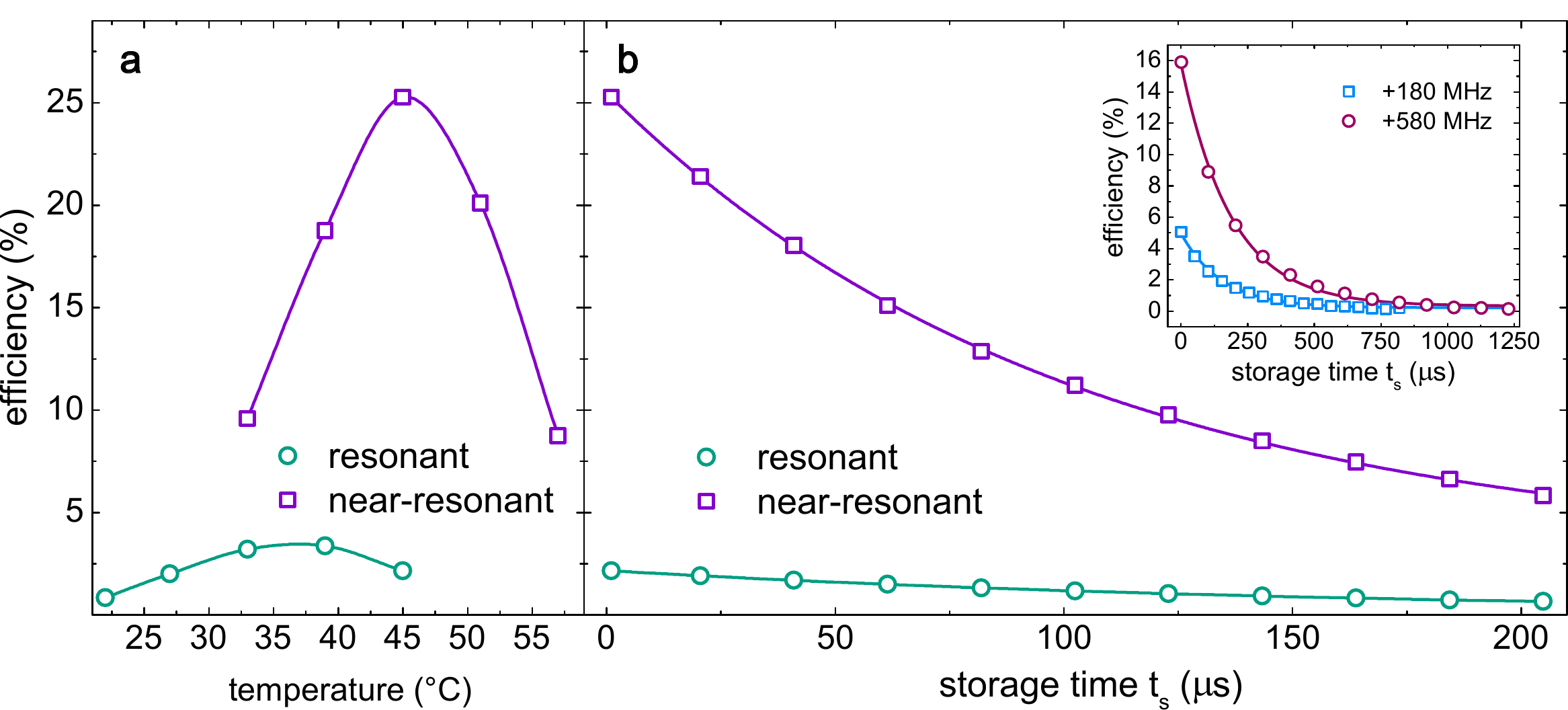}
\caption{(a) Comparison of optical memory efficiency for resonant (green symbols) and near-resonant with $\Delta_c=$-700 MHz (violet symbols) EIT schemes as a function of Rb vapor temperature. Higher efficiency is obtained in the near-resonant case, with the maximum efficiency reached at the temperature higher than the temperature of the resonant maximum efficiency. The lines are guide to the eye. (b) Optical memory efficiency as a function of storage time in the resonant (green symbols) and near-resonant case (violet symbols) at $\Delta_c=$-700 MHz, at Rb cell temperature of $\sim$45$^{o}$C. Inset depicts memory efficiency for $\Delta_c=$+180 MHz (blue symbols) and +580 MHz (bordeaux symbols). Lines are fit of the exponential decay function.)
	}
	\label{Fig7}
\end{figure}

As depicted in Fig.~\ref{Fig7}(a), there is an optimal temperature at which maximum memory efficiency can be achieved for a given detuning. 
This optimal temperature is determined by a compromise between three factors: the number of atoms participating in the coherent state, absorption and other decoherence mechanisms.
The memory efficiency initially increases as temperature is increased, due to the increased number of atoms that can be prepared in a coherent state.
However, an increase in temperature causes larger absorption of optical fields during propagation through the atomic vapor, which has a detrimental effect on memory efficiency \cite{ma2017}.
For example, absorption causes the reduction in the intensity of the coupling beam as it propagates through the atomic vapor. 
Consequently, this decrease in intensity results in a reduction in the contrast and width of the EIT resonance, thereby compromising the efficiency of storing the probe pulse in the vapor.
Moreover, other decoherence mechanisms, such as radiation trapping, spin-exchange collisions, stimulated Raman scattering, and four-wave mixing, \cite{nathaniel2008}, become prevalent, causing efficiency to drop with a further increase of temperature.

For a given temperature, higher memory efficiencies are achieved when the coupling laser is detuned from the $|5S_{1/2};F=3\rangle - |5P_{1/2};F'=3\rangle$ transition, i.e., by employing a near-resonant EIT scheme.
Detuning the coupling laser from resonance results in reduced absorption for the same number of atoms, leading to an increase in memory efficiency.
We achieve an approximately sixfold enhancement in memory efficiency when the near-resonant EIT $\Lambda$-scheme with $\Delta_c=$-700 MHz is used instead of a resonant one, as indicated by the green (resonant EIT) and violet (near-resonsnt EIT) symbols in Fig.~\ref{Fig7}(a).
Upon further increase of the detuning of the coupling laser the memory efficiency drops, and falls to zero for detunings beyond the Doppler broadened single-photon linewidth.  
The observed sweet spot in memory efficiency is observed for the coupling laser detuning at which the EIT feature fully transforms into a two-photon Raman resonance \cite{thomas2017}, see Fig.~\ref{Fig3} and the discussion related to it.

In Fig.~\ref{Fig7}(b) an optical memory efficiency as a function of storage time in the resonant (green symbols), and near-resonant (violet symbols) with $\Delta_c=$-700 MHz EIT scheme, at Rb cell temperature of $\sim$45$^{o}$C is shown.
Inset depicts measurement of the memory efficiency for the storage times up to 1.2 ms in the near-resonant EIT scheme ($\Delta_c=$+180 MHz (blue symbols) and $\Delta_c=$+580 (bordeaux symbols), at a Rb cell temperature of $\sim$45$^{o}$C.  
The decrease in maximum efficiency compared to $\Delta_c=$-700 MHz is attributed to increased absorption at smaller detunings of the coupling laser.

Lines in  Fig.~\ref{Fig7}(b) are fit of the exponential decay function $e^{-t/\tau}$ to the measured data, where $\tau$ denotes the memory lifetime.
It is worth noting that the increase in memory efficiency is not accompanied by an increase in memory lifetime. 
For a given vapor temperature, similar memory lifetimes are observed for both resonant and near-resonant EIT memory. 
For instance, in Fig.~\ref{Fig7}(b), we show the measured memory efficiency as a function of storage time, from which we extract memory lifetimes of $\tau = (105 \pm 3) \mu$s and $\tau = (101 \pm 3) \mu$s for resonant and near-resonant EIT, respectively.
In addition, for a given temperature, a similar memory efficiency is observed when the coupling laser is red- or blue- detuned from the atomic resonance, which is reasonable considering that amplitudes of EIT and two-photon absorption resonances shown in Fig.~\ref{Fig3}, are comparable for a given $|\Delta_{c}|$.

Finally, we find the optimal memory performance, with $25\%$ efficiency and storage times of up to 1.2 ms, in the near-resonant EIT scheme with $\Delta_c=$ - 700 MHz at a vapor temperature of $\approx 45~^{o}$C.

\section*{Conclusion}
\setcurrentname{Conclusion}
\label{Conclusion}

In summary, our study provides a comparison of light storage in antirelaxation-coated and buffer-gas-filled warm rubidium vapor cells using electromagnetically induced transparency (EIT). 
The buffer-gas-filled cell demonstrates a tenfold improvement in both storage time and memory efficiency compared to the antirelaxation-coated cells under identical experimental conditions. 
Additionally, the adoption of a near-resonant EIT $\Lambda$-scheme, as opposed to a resonant one, results in approximately a sixfold enhancement in memory efficiency while maintaining a comparable memory lifetime.
Through the optimization of buffer-gas-filled memory performance, we achieve a memory efficiency of 25$\%$, with storage times extending up to 1.2 milliseconds.
Together with our theoretical model, our findings contribute valuable insights to the field and provide novel perspectives that could optimize the development of high-performance, field-deployable quantum memories.

\section*{Methods}
\setcurrentname{Methods}
\label{methods}

\subsection*{Numerical model}

To theoretically investigate the EIT signals for a realistic $\Lambda$ system, we employ standard density matrix formalism in a four-level $^{85}$Rb atom interacting with two continuous-wave (cw) laser fields (coupling - $E_c$ and probe - $E_p$), as depicted in Fig.~\ref{Fig1}(a). 
The Hamiltonian of the system, denoted as $\hat{H} = \hat{H}_0 + \hat{H}_{int}$, comprises two components: the free atom Hamiltonian $\hat{H}_0$ and the interaction Hamiltonian $\hat{H}_{int}$. The latter describes the interaction of the atom with the two laser fields, specifically, $\hat{H}_{int}=\hat{H}_{14}+\hat{H}_{24}$. 
In the dipole approximation, parts of the interaction Hamiltonian can be expressed as $\hat{H}_{1n}=-\mu_{1n}\mathcal E_p(t)$ and $\hat{H}_{2n}=-\mu_{2n}\mathcal E_c(t)$, where $n=3,4$ and $\mu_{1n}$ and $\mu_{2n}$ denote the transition dipole moments of the relevant transitions, numbered according to Fig.~\ref{Fig1}(a) and calculated from Ref. \cite{Axner2004}.

Temporal evolution of the system is given by the density matrix equations of motion \cite{Boyd2008}:
\begin{eqnarray}
	\frac{\partial\rho_{nm}}{\partial t} & = & \frac{-i}{\hbar}\left[\widehat H,\widehat\rho\right]_{nm}-\gamma_{nm}\rho_{nm}, \,\,\,\,\,\ (n\neq m), \nonumber\\
	\frac{\partial\rho_{nn}}{\partial t} & = & \frac{-i}{\hbar}\left[\widehat H,\widehat\rho\right]_{nn}-\sum_{m(E_m<E_n)}\Gamma_{mn}\rho_{nn} \nonumber\\
	& & +\sum_{m(E_m>E_n)}\Gamma_{nm}\rho_{mm},
	\label{dens1}
\end{eqnarray}

where the subscripts $nm$ refer to the hyperfine levels numbered from the lowest to highest energy level (see Fig.~\ref{Fig1}). $\Gamma_{nm}$ gives the population decay rate from level $m$ to level $n$, while $\gamma_{nm}$ is the damping rate of the $\rho_{nm}$ coherence given by
\begin{equation}
	\gamma_{nm}=\frac{1}{2}(\Gamma_n+\Gamma_m)+\gamma_{nm}^{coll}.
	\label{gamakoh}
\end{equation}
Here $\Gamma_n$ and $\Gamma_m$ denote the total population decay rates of level $n$ and $m$. 
In our system $\Gamma_1 = \Gamma_2 = 0$, and $\Gamma_3 = \Gamma_4 = 2\pi \cdot 5.75$~MHz \cite{steck2}. 
$\Gamma_{nm}$ are calculated from $\Gamma_n$ and $\Gamma_m$ following Fermi's golden rule \cite{Axner2004}. 
The $\gamma_{nm}^{coll}$ depend on the particular alkali cell used.   
The $\gamma_{nm}^{coll}$ of optical transitions, i.e., $\gamma_{13}^{coll} = \gamma_{23}^{coll} = \gamma_{14}^{coll} = \gamma_{24}^{coll}$, arise from Doppler broadening, $2\pi~\cdot$~500~MHz, for anti-relaxation coated and Ne-filled cells, and from the Rb collisions with the Ne gas, $2\pi~\cdot$~25~MHz, for cell filled with 5 Torr of Ne gas \cite{derose2023}.
The $\gamma_{nm}^{coll}$ of non-allowed transitions, i.e. $\gamma_{12}^{coll}$ = $\gamma_{34}^{coll}$ are decoherence rates of ground and excited hyperfine levels, respectively, and their values are taken from measurements shown in Fig.~\ref{Fig5}(a), i.e. as b/2 in accordance with \cite{filkenstein2023, derose2023}.  

The system of Eqs.~(\ref{dens1}) is solved by invoking the rotating-wave approximation and introducing the laser electric fields $\mathcal E_c(t)=E_ce^{-i\omega_ct}$, and $\mathcal E_p(t)=E_pe^{-i\omega_pt}$, together with the slowly varying coherences $\sigma_{2n}=\rho_{2n}e^{-i\omega_ct}$ and $\sigma_{1n}=\rho_{1n}e^{-i\omega_pt}$, where $n=3,4$.
Here, $\omega_c$ and $\omega_p$ are coupling and probe laser frequencies defined as $\omega_c=\omega_{24}+\Delta_c$, $\omega_p=\omega_{14}+\Delta_p$, where $\omega_{24}$ and $\omega_{14}$ correspond to the $|5S_{1/2};F=3\rangle - |5P_{1/2};F'=3\rangle$ and $|5S_{1/2};F=2\rangle - |5P_{1/2};F'=3\rangle$ transitions, respectively.  
$E_c$ and $E_p$ are the amplitudes of the coupling and probe laser fields.       
Stationary solutions for atomic populations and coherences are obtained from the set of coupled differential equations, Eqs. (\ref{dens1}).
The probe absorption coefficient is proportional to $\alpha_{p}=-Im(\sigma_{14})$. 

\section*{Acknowledgements}

The authors acknowledge support from the Croatian Science Foundation (Slovenia-Croatia bilateral research project "Development of building blocks for new European quantum communication network" - IPS-2020-01-2616 and project "Frequency comb cooling of atoms" - IP-2018-01-9047).
In addition, this work was supported by the project Centre for Advanced Laser Techniques (CALT), co-funded by the European Union through the European Regional Development Fund under the Competitiveness and Cohesion Operational Programme (Grant No. KK.01.1.1.05.0001).
The authors acknowledge Vjekoslav Vuli\'{c} for his contribution in building a filtering cavity. 

\section*{Author contributions statement}

M.Đ. and D.B. conducted the experiment and analyzed the data, T.B. conceived the experiment, T.B. and N.Š. wrote the manuscript. N.Š., D.A. and T.B coordinated the experiment. All authors reviewed the manuscript.

\section*{Competing interests} The author(s) declare no competing interests.
\section*{Data availability} The datasets generated during and/or analysed during the current study are available from the corresponding author on reasonable request.

\end{document}